\documentclass{article}
\usepackage{graphicx} 
\usepackage{fullpage}
\usepackage{graphicx} 
\usepackage{cite}
\usepackage{amsmath,amssymb,amsfonts}
\usepackage{algorithmic}
\usepackage{textcomp,multirow}
\usepackage{cases,color,xcolor}
\usepackage[normalem]{ulem,mathtools}
\usepackage{subfigure,flushend}
\usepackage{adjustbox}
\usepackage[utf8]{inputenc}
\usepackage{dsfont}
\usepackage{multirow,array,booktabs}
\usepackage{comment}
\usepackage{multicol}
\usepackage[most]{tcolorbox}
\usepackage{xcolor}
\usepackage{soul}
\usepackage{titlesec}
\usepackage[super,sort&compress,numbers]{natbib}
\usepackage{authblk}

\newcommand{\Vl}{{V\hspace{-.3mm}l}}
\newcommand{\PVO}{{P\hspace{-.25mm}V\hspace{-.5mm}O}}
\newcommand{\ILsix}{{I\hspace{-.4mm}L_6}}

\title{\textbf{{A} novel mathematical model {for} predicting the benefits of physical activity {on} type 2 diabetes {progression}}}
 
\author{
    Pierluigi Francesco De Paola$^{1,2,3}$, Alessandro Borri$^{*,3}$, Fabrizio Dabbene$^{2}$,\\
    Karim Keshavjee$^{4}$,
    Pasquale Palumbo$^{5}$,
    Alessia Paglialonga$^{2}$
}

\affil{
\small 
    $^1$Politecnico di Bari, Bari, Italy\\
    $^2$Consiglio Nazionale delle Ricerche, Istituto di Elettronica e di Ingegneria dell’Informazione e delle
Telecomunicazioni (CNR-IEIIT), Turin, Italy\\
    $^3$Consiglio Nazionale delle Ricerche, Istituto di  Analisi dei Sistemi ed Informatica (CNR-IASI), Rome, Italy\\
    $^4$University of Toronto, Dalla Lana School of Public Health, Institute of Health Policy, Management and Evaluation, Toronto, Canada\\
    $^5$University of Milano-Bicocca, Department of Biotechnologies and Biosciences, Milan, Italy\\
    $^*$Corresponding author: alessandro.borri@iasi.cnr.it
}

\makeatletter
\renewcommand\@biblabel[1]{#1 }
\makeatother

\titleformat{\section}
{\color{black}\normalfont\Large\bfseries}{\thesection}{1em}{}

\begin{document}
\date{}

\maketitle

\section*{Abstract}
Despite the well-acknowledged benefits of physical activity for type 2 diabetes (T2D) prevention, the literature surprisingly lacks validated models able to predict the long-term benefits of exercise on 
{T2D} progression and support personalized risk prediction and prevention.  
To bridge this gap, we developed a novel mathematical model that formalizes the link between exercise and {short- and long-term }glucose-insulin dynamics 
to predict the {benefits} of regular exercise on T2D progression. 
The model quantitatively captured the dose-response relationship (larger benefits with increasing intensity and/or duration of exercise 
), it consistently reproduced the benefits of clinical guidelines for diabetes prevention,
and it accurately predicted persistent benefits {following} 
interruption of {physical activity},  
{in line} with real-world {evidence from} the literature. {These} results are encouraging and can be the basis for future development of decision support tools able to assist patients and clinicians in tailoring preventive lifestyle interventions. 
\section*{Introduction}
Type 2 diabetes (T2D) is a high-prevalence, high-burden chronic disease, whose impact is becoming more and more concerning for public health worldwide 
\cite{Khan2019}, 
with the global prevalence projected to increase up to 592 millions by 2035\cite{guariguata2014global} and {with} severe economic implications {for the} healthcare systems \cite{bommer2017global,ZimmetEtAl2001}. However, {evidence suggests that }T2D can be prevented, for example through
lifestyle interventions including physical activity\cite{ZimmetEtAl2001,knowler2002diabetes,eriksson1991prevention,lindstrom2006sustained,kosaka2005prevention,diabetes2002reduction,hemmingsen2017diet}. 
{Digital decision support tools incorporating m}athematical models can be helpful to {quantitatively} predict the benefits of lifestyle interventions on T2D progression {and can} provide a basis for {reducing the individual risk through tailored recommendations}
for T2D prevention. {For example, predictive models able to estimate the risk of T2D as a function of individual health status and lifestyle on a daily basis may be integrated into digital patient monitoring tools to continuously track individual biomarkers and support patients in achieving targets and reduce their risk through behavior change\cite{zahedani2023digital, salunkhe2023digital}.}
Several mathematical models were introduced in the literature to describe the slow dynamics of {T2D} progression in the long term {as a function of, e.g., individual blood glucose levels, insulin concentration, or mass of pancreatic beta-cells}\cite{ToppEtAl2000,HaEtAl2016,de2008mathematical,de2019novel}.
These models are well-established and show promise for predicting the risk of developing T2D along a time span 
of several years. However, they do not include mechanistic descriptions of the beneficial effects of {exogenous interventions, for example} physical activity, on T2D risk. Regarding the effects of physical activity, {most of} the models introduced so far are focused on the short-term effects 
of a single exercise session, on a timescale of minutes to hours \cite{al2021glucose,palumbo2018personalizing,roy2007dynamic,derouich2002effect}. 
Hence, the {available} models {of the short-term effects of physical activity} are suitable to be employed in short-range closed-loop glucose control techniques, whose effectiveness has been shown in several works \cite{tsoukas2021fully, kovatchev2020evening}. However, these short-term models are not able to quantitatively describe the cumulative beneficial effects of repeated exercise sessions on the slow dynamics of T2D progression - effects that are known to occur on a timescale of weeks, months, or even years\cite{PetersenPedersen2005,CurranEtAl2020,diabetes2002reduction}.
The aim of this study is to develop a novel \textit{two-timescale} model incorporating {both} the short-{term} and {the} long-term {cumulative} effects of physical activity on T2D progression and {to assess the ability of the proposed model} to describe real-world evidence on the benefits of different programs of physical activity, as available from the literature {and from preventive recommendations} 
\cite{bull2020world,boonpor2023dose,lindstrom2006sustained,li2008long}.
In our previous work \cite{DePaolaEtAl}, 
we developed a preliminary version of a two-timescale model by combining a well-established model of long-term T2D progression
\cite{HaEtAl2016} with (i) a model of the short-term effects of individual sessions of physical activity
\cite{roy2007dynamic}
and (ii) a newly developed model of the long-term, cumulative effects of repeated sessions of physical activity 
mediated by Interleukin-6 (IL-6), a protein with anti-inflammatory action released by muscle tissue under exercise \cite{MorettiniEtAl2017} that contributes in preserving beta-cell mass and function
\cite{PetersenPedersen2005,CurranEtAl2020,MorettiniEtAl2017}.
Specifically, we modeled the release of IL-6 induced by physical exercise 
as suggested by Morettini et al.\cite{MorettiniEtAl2017} and we {described} the integral, cumulative effect of IL-6 released during repeated exercise sessions {by introducing} an additional state variable {that} represents the total volume of IL-6 {and} acts on the balance between beta-cell proliferation and apoptosis, therefore regulating insulin release and glucose metabolism.
Notably, the preliminary model proposed in our previous work \cite{DePaolaEtAl} is the first one  accounting for the different time scales that involve physical activity (minutes) and T2D progression (years) by combining a short-term and a long-term dynamics, and was able to
consistently predict an increase in beta-cell mass (and, therefore, an increase in insulin {concentration}) following regular physical activity, leading to a decrease in basal glucose levels and therefore T2D prevention. However, the 
model {did not account for} the effects of varying intensity and duration of physical activity {on beta-cell proliferation and apoptosis and did not include the benefits of physical activity on individual insulin sensitivity}.
In pursuit of the ultimate goal of developing and validating a digital support tool capable of accurately predicting the benefits of physical activity for T2D risk monitoring and prevention, this study has a dual objective:
(i) {to model the influence of varying physical activity on T2D progression,} we modify and finely tune our preliminary model \cite{DePaolaEtAl} {by} improving its capability to capture the long-term 
effects of IL-6 on beta-cells dynamics and on insulin sensitivity; 
(ii) {to investigate if, and to what extent, the model predictions are aligned with real-world evidence}, we 
assess the predictions of the model {by considering} different time courses {of T2D} progression {(i.e., different time constants of the simulated decay in insulin sensitivity)} in a range of {simulated} conditions, specifically: different exercise programs at varying intensity and {weekly} duration, {different} World Health Organization (WHO) recommendations {for chronic disease prevention}, and
de-training {in different diabetes prevention programs}.

\section*{Results} 

\begin{figure*}[]
\begin{tcolorbox}[colback=white, colframe=white, sharp corners, boxrule=0.5pt]
{\includegraphics[width=\textwidth]{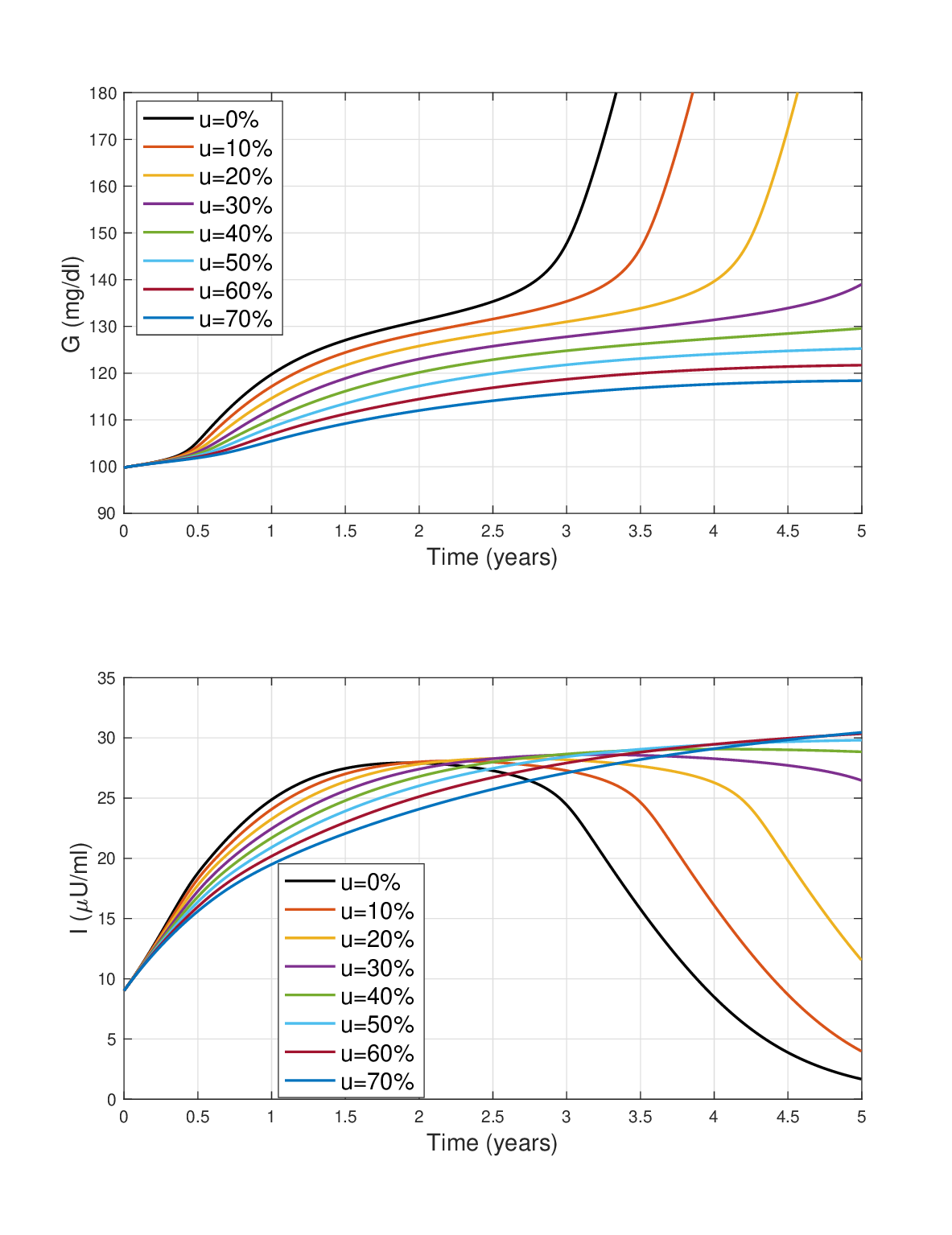}}
\end{tcolorbox}
\caption{
{Basal} glucose concentration (top panel) and insulin concentration (bottom panel) {as a function of $u$} over a five-year simulation horizon {(simulations: 60 minutes/session, 3 sessions/week, $\tau_{SI}=150$ days).}}
\label{fig:DoseResponse}
\end{figure*}

\begin{figure*}[]
\begin{tcolorbox}[colback=white, colframe=white, sharp corners, boxrule=0.5pt]
{\includegraphics[width=\textwidth]{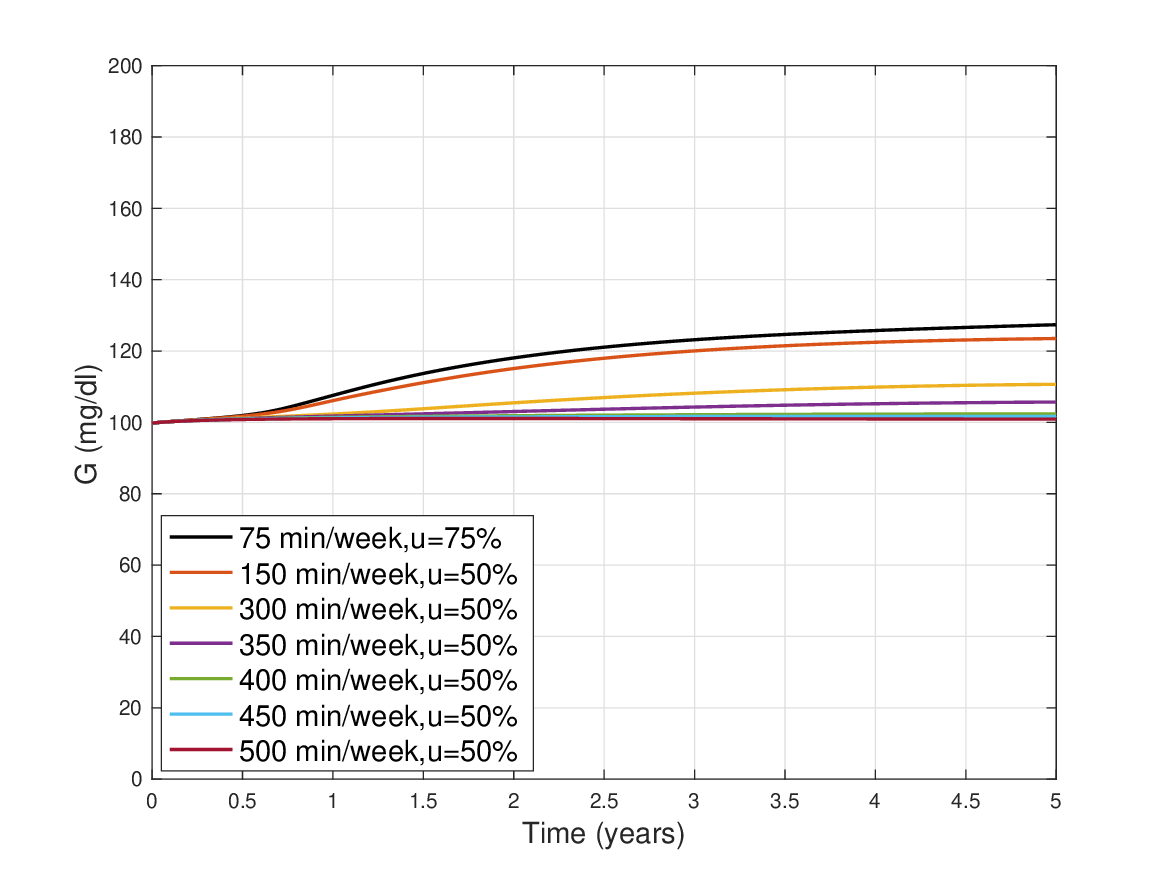}}
\end{tcolorbox}
\caption{{Basal} glucose concentration 
{of vigorous exercise ($u=75\%$) for 75 minutes/week and moderate exercise ($u=50\%$) as a function of exercise duration, from 150 to 500 minutes/week, using $\tau_{SI}=180$.}} 
\label{fig:ExercisePrograms}
\end{figure*}

\subsection*{{Simulated effects of varying exercise intensity}}
{The predictions of the proposed model obtained by simulating different exercise programs with varying intensity are shown in Fig.~\ref{fig:DoseResponse} and Table \ref{tab:GlucoseTrend1}.}  Fig.~\ref{fig:DoseResponse} shows the basal glucose concentration {($G$)} and the insulin concentration {($I$)} 
as a function of the exercise intensity ($u$) (from $0$ to $70\%$) 
obtained by simulating a {training} program of 3 sessions/week, 60 minutes/session in a time window of 5 years,
with {time constant of insulin sensitivity decay (}$\tau_{SI}$) equal to $150$ days (thus assuming a {fast} diabetes course). 
Fig.~\ref{fig:DoseResponse} shows that 
the benefit produced by exercise, as determined by lower values of {$G$} and higher values of {$I$}, increases with increasing $u$, in line with the well-known dose-response relationship between physical activity and benefit 
\cite{boonpor2023dose}.
In particular, 
simulations show that for $u\geq50\%$ {(i.e., moderate-to-vigorous intensity)}, {$G$} 
does not show the steep inflection that is observed 
for $u<50\%$ {and for no physical activity ($u=0\%$)}. Moreover, for $u=60\%$ and $u=70\%$,
after five years $G$ remains below the diabetic threshold of 126 mg/dl.
Table \ref{tab:GlucoseTrend1} shows the values of $G$ and $I$ observed over a longer time horizon (at $t=$ 10, 15, and 20 years) for {$u=50\%$, $60\%$, and $70\%$ (i.e., moderate-to-vigorous intensity) and for $\tau_{SI}=150$ and $180$ days (i.e., faster and slower disease course, respectively)}. 
For $\tau_{SI}=150$ days {and $u=50\%$}, $G$ reaches 
the diabetic threshold 
after ten years, and then it steeply increases. 
For higher values of $u$, 
no inflection is observed and $G$ remains below the diabetic threshold {in the whole simulation window} and reaches values close to the healthy range {(i.e., down to about 100 mg/dl)}. 
When a slower course of T2D is assumed (i.e., $\tau_{SI}=180$), 
no inflection {in $G$} is observed with $u\geq50\%$, {$G$} remains below the diabetic threshold and gets close to the healthy range after 20 years. 
Overall, these results suggest that T2D onset could be delayed with a regular physical activity program of three sessions/week, 60 minutes/session at moderate or vigorous intensity, and that increased benefit is observed with increasing intensity, in line with {the well-known dose-response relationships reported in} the literature \cite{bull2020world,boonpor2023dose}.
For what concerns basal insulin concentration ($I$), 
{Fig. \ref{fig:DoseResponse} shows that} it reaches progressively higher values at $t=5$ years as $u$ increases, due to the increased
beta-cell mass promoted by exercise.
Over the 20-year {horizon},
when exercise is not sufficient to prevent T2D ($\tau_{SI}=150$ days, $u=50\%$), $I$ progressively drops due to decreased beta-cell mass. Instead, when vigorous intensity exercise {is simulated} ($u=60\%$, $u=70\%$),
$I$ increases and, thus, $G$ decreases {down to values close to the healthy range}. {Specifically,} $I$ reaches a steady state condition around 40 $\mu$U/ml, whenever a basal glycemia of $100$ mg/dl is restored.
\subsection*{{Simulated effects of WHO recommendations}}
Fig.~\ref{fig:ExercisePrograms} shows the {results }observed {by simulating} the WHO preventive recommendations {for cardiovascular and chronic disease prevention} \cite{bull2020world} {on a five-year horizon. Specifically, Fig. \ref{fig:ExercisePrograms} shows the trend in $G$ observed by simulating vigorous-intensity physical activity for 75 minutes/week and moderate-intensity physical activity for 150 to 500 minutes/week.} 
The curves {of the two minimum recommended exercise programs (i.e., vigorous-intensity physical activity for 75 minutes/week and moderate-intensity physical activity for 150 minutes/week)} remain very close {to each other} for the whole duration of the simulation, being the difference between the curves at the fifth year around {only} 4 mg/dl, in line with the equivalence of the two training programs reported by the WHO \cite{bull2020world}. It should be noted that the small difference between the two curves is {negligible and may be} further {reduced} if slightly different values of $u$ are used ({i.e,} higher values of $u$ for moderate physical activity, lower values of $u$ for vigorous physical activity).
{I}f the weekly duration of moderate-intensity exercise is increased up to 500 minutes/week{, $G$ further decreases and the observed benefit almost saturates} at weekly duration of 
400 minutes or higher, {fully} in line with the findings by  Boonpor et al. \cite{boonpor2023dose}. 
\subsection*{{Simulated effects of de-training in diabetes prevention studies}}
Table \ref{tab:CombinedTable} presents the values of $G$ obtained by simulating the Finnish Diabetes Prevention Study (FDPS) and the China Da Qing Diabetes Prevention Study (CDQDPS) programs,
which include regular physical activity programs followed by a discontinuation of exercise after four and six years, respectively. As for FDPS, following  the interruption of the intervention at the fourth year, the simulated progression of T2D is significantly delayed at the seventh year in three out of the six simulated cases, specifically in those cases where $G$ was {already} below the diabetic threshold at the end of the intervention {(i.e., $u=40\%$, $\tau_{SI}=180$ days; $u=50\%$, $\tau_{SI}=150$ and $180$ days)}. 
In the other three cases, 
the values of $G$ were in the diabetic range at {the end of the training at} the fourth year, and simulations suggest 
an irreversible disease progression at the seventh year. simulation.
Similar results are observed with the simulated CDQDPS. Specifically, an irreversible increase in $G$ is observed {after discontinuation of exercise} at $u=30\%$, i.e., in those cases where $G$ is already close to or in the  diabetic {range} at the sixth year.
In three cases ($u=40\%$, $\tau_{SI}=180$ days;  $u=50\%$, $\tau_{SI}=150$ and $180$ days), a complete reversal is observed, with $G$ restoring to a normoglycemic condition at about 100 mg/dl {at the end of the simulation window}, whereas for $u=40\%$ and $\tau_{SI}=150$ days a significant delay is observed after 20 years, {with} 
$G$ {remaining almost stable and slightly below} the diabetic threshold. 
{In general, {for what concerns the effects of de-training following a regular exercise program,} Table \ref{tab:CombinedTable} shows that (i) the higher the intensity $u$ and the slower the decay {of insulin sensitivity}, the lower the values of $G$ at the end of the intervention and at the end of the follow-up period, and (ii) when $G$ is lower than the diabetic threshold at the end of the intervention, benefits can be maintained in the follow-up period, up to 14 years after the discontinuation.}

\section*{Discussion}
The  role played by physical activity in delaying or preventing of T2D makes exercise a great ally in the battle against this serious disease, which is forecast to represent a huge challenge worldwide \cite{Khan2019}. What makes exercise \textit{crucial} {in T2D prevention} is its {beneficial} effect that contributes to preserve beta-cell function, which is actually {one of} the key factor{s} for diabetes prevention \cite{PetersenPedersen2005,CurranEtAl2020}. {Digital tools incorporating mathematical models can be of great help to support patients in reaching their healthy lifestyle targets and prevent T2D}\cite{greenwood2017systematic,so2018telehealth}. 
Nevertheless, the literature shows {a} lack of mathematical models able 
to suggest targeted recommendations for {physical activity} intervention to individuals at risk of developing T2D. 
The model presented in this study is the first in the literature able to describe the short- and long-term effects of physical activity on T2D progression, particularly the cumulative benefits on beta-cell function {and on insulin sensitivity} mediated by IL-6 released during exercise.  
The proposed model predicts the trends of glucose and insulin concentration {in individuals in} the early phases of {T2D} progression undergoing regular physical exercise. {The model is }simulated under a range of conditions, and predictions are validated in relation to findings reported in the literature.
For example, the {modeled} increase (or the slowdown in the decrease) of insulin sensitivity {due to physical activity} 
({as shown in the }Methods {section, }Fig.~\ref{fig:SI_colors})
is aligned with findings reported in the work by Damaso et al.\cite{damaso2014aerobic} and Uusitupa et al. \cite{uusitupa2003long} for moderate physical activity, specifically an improvement of about 29\% at the first year and about 20\% at the the fourth year, respectively. Moreover, the observed
vanishing of the improvement in the long-term 
is consistent with what discussed in Balducci et al.\cite{balducci2022sustained}. {Moreover, the proposed model is sensitive to changes in exercise intensity, as shown in Fig. \ref{fig:DoseResponse} and Fig. \ref{fig:subplotG-beta} (Methods section). Specifically,}
the model accurately captures the dose-response relationship \cite{boonpor2023dose, balducci2022sustained, gregg2012association} between physical activity and long-term benefit {on disease progression}. For example, {the model predictions show that} 
the risk of developing T2D decreases as the exercise intensity increases, all the other parameters (weekly duration, $\tau_{SI}$, initial conditions) being equal (Fig.~\ref{fig:DoseResponse}, Tables \ref{tab:GlucoseTrend1}-\ref{tab:CombinedTable}).
Similarly, the same benefit {is predicted} with moderate intensity exercise for 150 minutes/week and with vigorous-intensity {exercise} for 75 minutes/week (Fig. \ref{fig:ExercisePrograms}), in line with the WHO \cite{bull2020world}, {the} American Diabetes Association\cite{colberg2016physical} and {the} American College of Sports Medicine \cite{haskell2007physical} recommendations. 
In addition, Boonpor et al.\cite{boonpor2023dose} state that usual recommendations on physical activity ({e.g.,} around 150 minutes/week, moderate exercise) can delay diabetes up to about 13 years (as shown in Table \ref{tab:GlucoseTrend1}) and that increasing the dose {(e.g., the weekly duration)} beyond 
usual recommendations could lead to greater benefits (as shown in Fig. \ref{fig:ExercisePrograms}), up to a maximum observable benefit 
around 400 minutes/week {for moderate-intensity physical activity}, where 
the risk reduction levels off. Our results are {fully} aligned with this evidence. 
Moreover, the proposed model was also able to predict a maintained benefit following 
discontinuation of the intervention, as 
outlined in Table \ref{tab:CombinedTable}, where the findings from the FDPS and CDQDPS are {precisely} replicated. 
It is worth noting that in the FDPS \cite{uusitupa2003long} the average glycemia observed in the group of participants who developed T2D was 126 mg/dl at the fourth year. This value is accurately approximated by our model as the average value of $G$ predicted after four years in the conditions in which T2D eventually occurs {is equal to} 127 mg/dl (Table \ref{tab:CombinedTable}).
{Similar consistence was observed} for {the simulated} CDQDPS {training and de-training experiment.} 
{The} results {reported in Table \ref{tab:CombinedTable} }seem to capture the different {T2D} incidence scenarios {and glycemia levels} reported in the work by Li et al. \cite{li2008long}. In some cases, T2D develops at the sixth year or along the 20-year simulation, and {in some other} cases {T2D is substantially} slowed down or even reversed. 
Interestingly, Table \ref{tab:CombinedTable} shows that (i) the higher the intensity $u$ and the slower the drop in $S_{I}$, the lower the  predicted values of $G$ at the end of the intervention and at the end of the follow-up period; and (ii) when $G$ is lower than the diabetic threshold at the end of the intervention, benefits are maintained in the follow-up period, up to 14 years after the discontinuation if a moderate-intensity exercise program has been performed for six years.
The present work provides promising results {as the model developed proved} suitable to accurately describe the effects of physical activity {and provide predictions} consistent with a range of clinical findings. However, the study has some limitations that should be overcome in the future developments of the study.
In particular, 
some of the referred works involve a combined program of diet and exercise \cite{diabetes2002reduction,lindstrom2006sustained,li2008long,balducci2022sustained}and do not specifically report the 
separate contribution of physical activity. {It is to note that} research about the separate benefits of diet and exercise in lifestyle interventions for T2D prevention has not reached a consensus. However, it will be important to investigate the {separate} effects of exercise {and diet} to develop tailored recommendations for each component of the intervention. Further developments {of the model} should {also} include incorporation of diet and the balance between caloric intake and expenditure into the model.
Moreover, it will be important to investigate the benefits of physical activity in healthy individuals at low risk of T2D (for example, by introducing different modeling approaches for the decay of insulin sensitivity and by modeling different individual fitness levels) to address 
the \textit{key} role of exercise in  early {T2D 
 } prevention \cite{ToppEtAl2000,defronzo2011preservation,defronzo1988triumvirate}. 
These developments could pave the way to a more comprehensive personalization of model predictions in a range of scenarios to support the development of  
targeted recommendations for effective prevention of T2D. Future digital tools could incorporate these personalized models to simulate virtual subjects and precisely predict the benefits of exercise {in order to} identify {personalized} recommendations for reducing the risk of T2D according to the {individual characteristics and the monitored} progression of the disease.

\section*{Methods}
\subsection*{Model formulation}
 
\begin{figure*}[!ht]
 \begin{tcolorbox}[colback=white, colframe=white, sharp corners, boxrule=0.5pt]
{\includegraphics[width=\textwidth]{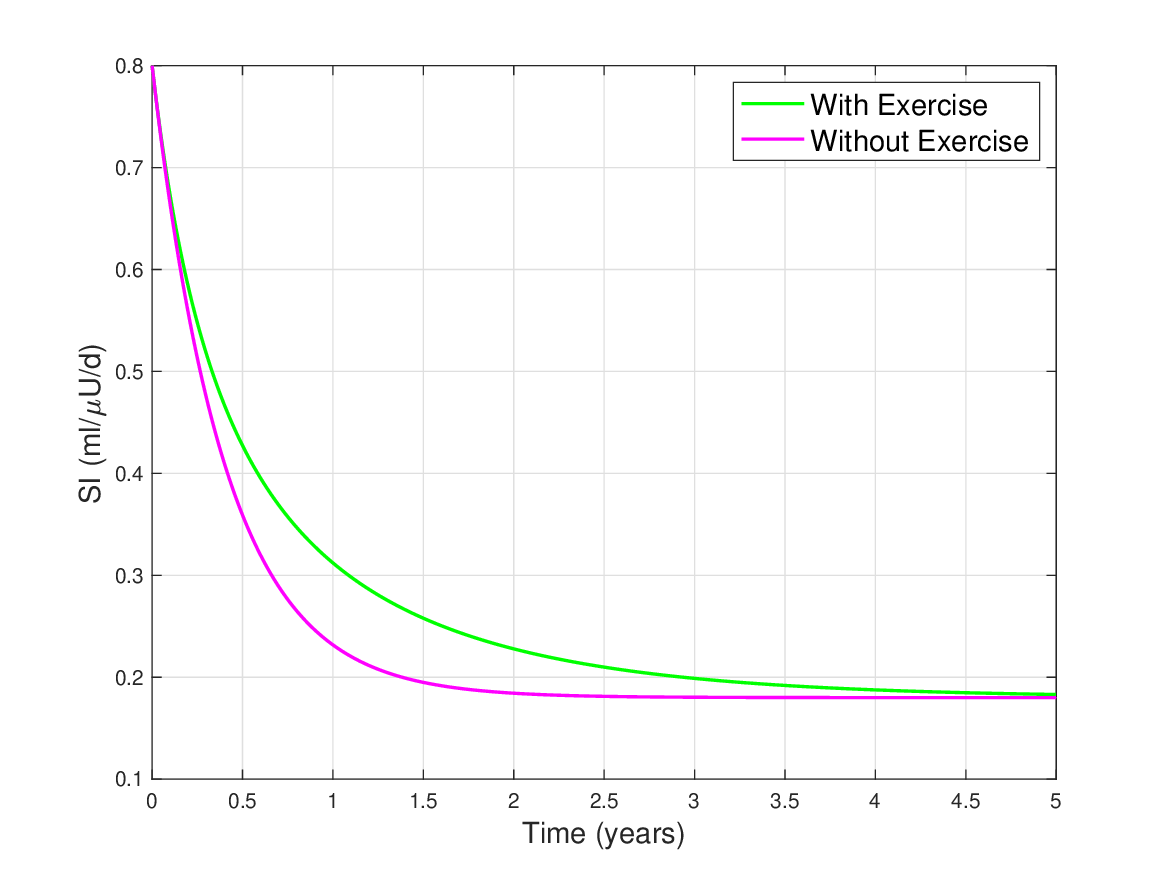}}
\end{tcolorbox}
\caption{{Time trend of insulin sensitivity over a five-year horizon {(simulation: 60
minutes/session, 3 sessions/week, $u = 50\%$, ${S_{I\!,\text{\it target}}=0.18}$, $\tau_{SI}=150$ days)}. The green curve {shows} the increase produced by the exercise with respect to the trend observed when no intervention is considered (magenta curve).}}

\label{fig:SI_colors}
\end{figure*}

The model proposed in this study is derived from, and improves upon, the preliminary one presented in detail in our previous work\cite{DePaolaEtAl}. The main equations of the proposed model are {reported in} the following:
\begin{subequations}
	\begin{align}
         &\bar{P}(ISR) = P(I\hspace{-.4mm}S\hspace{-.4mm}R) \cdot\left(1+ \zeta_1\frac{\Vl^{2}}{\ {k_n^2}+\Vl^{2}}\right),\label{eq:P_bar} \\
        &\bar{A}(M) = A(M)\cdot\left(1-\zeta_2\frac{\Vl^{2}}{{k_n^2}+\Vl^{2}}\right),\label{eq:A_bar} \\
         &\dot \beta = \frac{\bar{P}(I\hspace{-.4mm}S\hspace{-.4mm}R)-\bar{A}(M)}{\tau_{\beta}} \beta, \\
       &{\dot S_I  = \frac{-(S_I-S_{I\!,\text{\it target}})}{\tau_{SI}}}\cdot\left(1-\zeta_3\frac{\Vl}{\ {k_{n,si}}+\Vl}\right), \label{eq:SI} \\
        &\dot{\PVO}_2^{\max} = -0.8 \PVO_2^{\max} + 0.8 u,\label{eq:oxygen} \\
        &\dot{\ILsix} = SR \cdot \PVO_2^{\max} - K_{IL6} \cdot \ILsix \label{eq:ILsix}, \\
		&\dot{\Vl} = \ILsix - k_s \Vl, \label{eq:Vl}\\
		&\dot G = R_0+{\frac{W}{V_g}}\left(G_\text{\it prod}-G_\text{\it up}\right)-\left(E_{g0}+S_I I\right) G, \label{eq:G} \\
		&\dot I =\frac{\beta }{V}I\hspace{-.4mm}S\hspace{-.4mm}R - k I - I_e, \\ \label{eq:I} 
        &\dot \gamma  = \frac{\gamma_{\infty}(G)-\gamma}{\tau_{\gamma}}, \\
		&\dot \sigma  = \frac{\sigma_{\infty}(I\hspace{-.4mm}S\hspace{-.4mm}R,M)-\sigma}{\tau_{\sigma}},
		 \\	
	\end{align}

\end{subequations}
where $\zeta_1=10^{-4}$, $\zeta_2=10^{-4}$, $k_n=10^{6}$ ((pg/ml) min), $\zeta_3=1.4$, $k_{n,si}=5\cdot10^6$ ((pg/ml) min).
 For a complete description of variables and parameters reference is made to our previous work \cite{DePaolaEtAl} and to the work by Ha et al.\cite{HaEtAl2016}.
 Specifically, 
to improve the long-term dynamics of the model, in this study we 
reformulate the equations that describe the integral effect of the released IL-6 {($\ILsix$)} on the dynamics of beta-cells by introducing a multiplicative contribution in place of the additive contribution originally used in our preliminary work \cite{DePaolaEtAl} to scale the effect of the exercise with respect to natural beta-cell proliferation ($\bar{P}(ISR)$) and apoptosis ($\bar{A}(M)$) (equations \eqref{eq:P_bar}-\eqref{eq:A_bar}), where $ISR$ (insulin secretion rate) and $M$ (metabolic rate) are defined in the work by Ha et al.\cite{HaEtAl2016}. In addition, here we model for the first time the effect of physical activity 
on insulin sensitivity (equation \eqref{eq:SI}). 
Physical activity is captured through the state variable ${\PVO}_2^{\max}$, representing the \textit{sovrabasal} oxygen consumption during exercise as a function of the exercise intensity $u$ 
(equation \eqref{eq:oxygen}). The variable ${\PVO}_2^{\max}$, in turn, promotes the dynamics of  $G_{prod},G_{up}$ and $I_e$ (describing the short-term impairment of blood glucose regulation induced by the exercise as described in our previous work \cite{DePaolaEtAl}), and the dynamics of $\ILsix$, that is released by muscle tissues under exercise (short-term) and exerts its integral anti-inflammatory action (long-term), as expressed by the state variable $\Vl$ representing the cumulative effect of $\ILsix$ (equations \eqref{eq:ILsix}-\eqref{eq:Vl}). 
Specifically, $\Vl$
leads to an improvement of insulin sensitivity ($S_I$) and an increase of beta-cell mass ($\beta$) due to increased proliferation ($\bar{P}(ISR))$ and reduced apoptosis ($\bar{A}(M)$). As a result, 
the basal insulin concentration ($I$) increases (equation \eqref{eq:I}) and, thus, lower values of basal glucose concentration ($G$) can be obtained. 
In equation \eqref{eq:SI}, 
a Michaelis-Menten function of $\Vl$ is used to account for a larger effect of physical activity in the first year of the intervention, 
as suggested by Damaso et al. \cite{damaso2014aerobic} and Bird et al.\cite{bird2017update}, and the parameters $\zeta_3$ and $k_{n,si}$ were set 
to replicate the average improvement in $S_I$ {following one-year and four-year physical activity} reported in by Damaso et al.
\cite{damaso2014aerobic} 
and Uusitupa et al.\cite{uusitupa2003long}, respectively. 
An exemplary representation of the behavior of equation \eqref{eq:SI} with a simulated exercise program of 60 minutes/session, 3 sessions/week at $u= 50 \%$ (as in Damaso et al.\cite{damaso2014aerobic}) 
is shown in Fig.~\ref{fig:SI_colors}, where 
${S_{I\!,\text{\it target}}=0.18}$ and $\tau_{SI}=150$ days (i.e., simulating a rapid T2D progression). 
The example in Fig.~\ref{fig:SI_colors} shows that an improvement in $S_I$ {of} about 30\% is observed 
at the end of the first year, in line with findings shown by Damaso et al. \cite{damaso2014aerobic}, whereas after four years 
the distance between the two curves falls below 8\%.
Similarly, when the same exercise program is simulated by setting $\tau_{SI}=180$ {days}, (i.e., a slower trend for T2D progression), a higher benefit on $S_I$ is observed, {specifically} about 35\%  after one year and about 20\% after four year{s, paralleling results shown by Uusitupa et al. \cite{uusitupa2003long}}. 

\begin{figure*}[!ht]
\begin{tcolorbox}[colback=white, colframe=white, sharp corners, boxrule=0.5pt]
{\includegraphics[width=\textwidth]{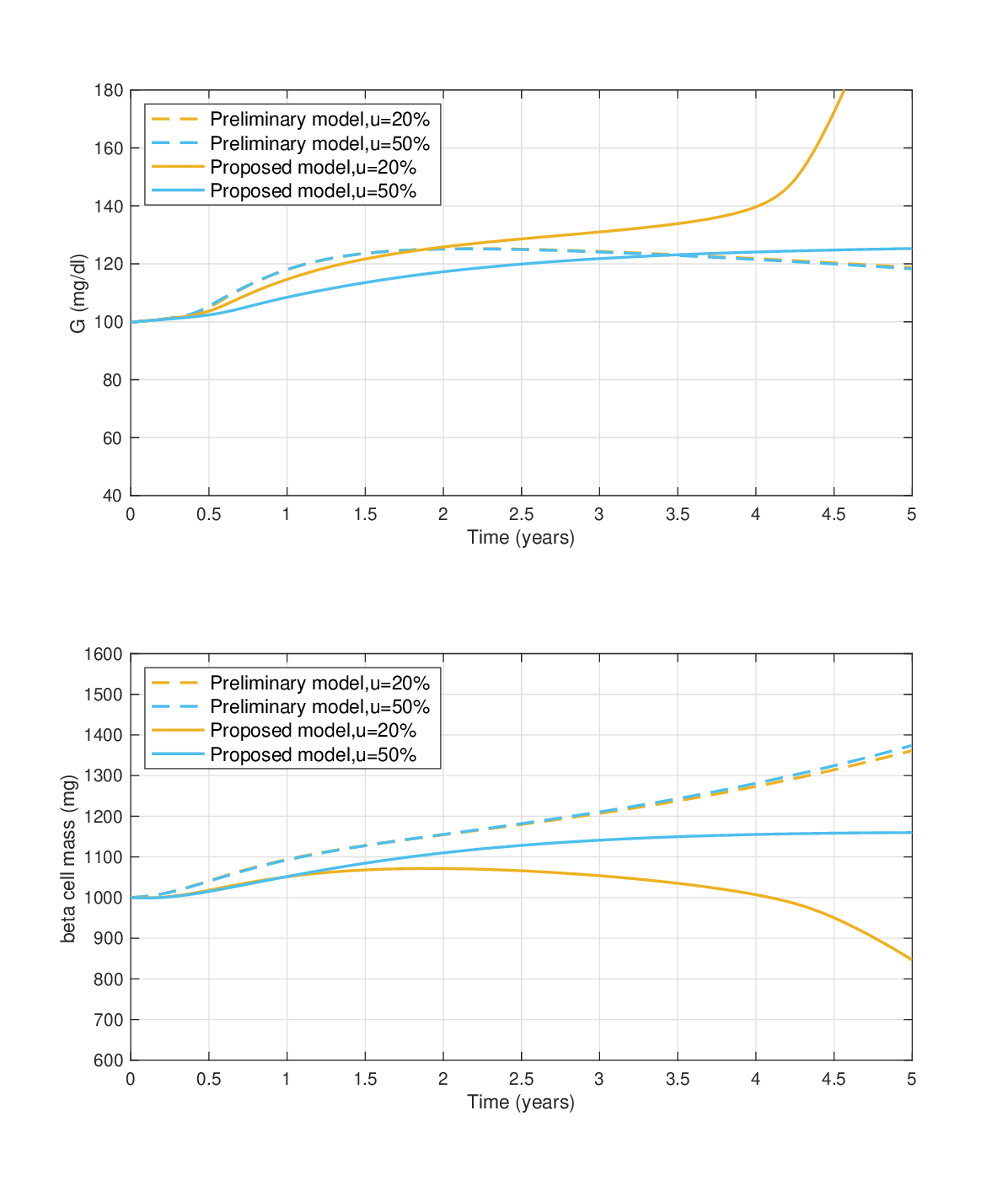}}
\end{tcolorbox}
\caption{Glucose concentration (top panel) and beta-cell mass (bottom panel) as a function of time {obtained using} the preliminary version of the model {\cite{DePaolaEtAl}} (dotted lines) and the model here proposed {(Equations }{(1)}, continuous lines) {(simulation: 60
minutes/session, 3 sessions/week, $u = 20\%$ and $50\%$, ${S_{I\!,\text{\it target}}=0.18}$, $\tau_{SI}=150$ days).}}

\label{fig:subplotG-beta}
\end{figure*}

In equations \eqref{eq:P_bar}-\eqref{eq:A_bar}, the gains of the Hill functions $\zeta_1$, $\zeta_2$ and $k_n$ are set with the aim of targeting a glycemia value around 126 mg/dl after 
five years of the same exercise program simulated in Fig. \ref{fig:SI_colors} and  given equation \eqref{eq:SI} described above, 
consistently with the results described in the Diabetes Prevention Program study\cite{diabetes2002reduction}. 
Fig. \ref{fig:subplotG-beta} shows a comparison between 
the preliminary version of the model \cite{DePaolaEtAl} and the model  proposed in this study, at two different exercise intensities.  
Specifically, Fig. \ref{fig:subplotG-beta} shows that the preliminary version {of the model led to} early saturation of the effects of physical activity on beta-cell dynamics. As such, the model was not able to 
capture the effects due to changes in duration or intensity of the exercise, leading to unrealistic 
reversal of diabetes for any exercise program. Conversely, the model here developed shows different benefits for varying intensities, in line with the dose-response relationship investigated by the simulations described in the following section. 
\subsection*{Model simulations}
The behavior of the model was characterized by simulating a range of physical activity programs, as reported in the literature and in well-known T2D preventive guidelines. Specifically, the predicted benefits of physical activity were characterized by addressing three key aspects, i.e.: (i) the effect of varying exercise intensity; (ii) the effect of applying the World Health Organization (WHO) recommendations for chronic disease prevention; (iii) the effect of a discontinued exercise program, as presented in more detail in the following.
All the simulations were carried out by supposing a drop in insulin sensitivity $S_{I}$ from $0.8$ (at $t=0$) to $0.18$ (at $t=5$ years) described by an exponential decay to simulate the onset of the conditions predisposing to T2D, following a typical approach in mathematical models of T2D progression\cite{ToppEtAl2000,HaEtAl2016}. The time constant of the exponential decay ($\tau_{SI}$) was set to two different values, i.e.: $\tau_{SI}=150$ days and $\tau_{SI}=180$ days,
to simulate different time courses
of T2D progression \cite{ToppEtAl2000,HaEtAl2016}. Initial conditions for the variables $G$ (mg/dl), $I$ $(\mu$U/ml), $\beta$ (mg), $\gamma$, and $\sigma$ ($\mu$U/$\mu$g/day) were $[99.7604,9.025, 1000.423,-0.00666,536.67]$ respectively, whereas $\PVO_2^{\max}$, $G_\text{\it prod}$ (mg/kg/min), $G_\text{\it up}$ (mg/kg/min), $I_\text{\it e}$ ($\mu$U/ml/min), $\ILsix$ (pg/ml), and $Vl$ ((pg/ml)min) were initialized to $0$, as in our previous work \cite{DePaolaEtAl}.

\subsubsection*{Simulations of a regular exercise program with varying intensity}
    The existence of a
    dose-response relationship between the intensity of physical activity and the incidence of T2D is well known from the literature
    \cite{boonpor2023dose,balducci2022sustained}.
    To assess the ability of the proposed model to predict the {increased} benefits of physical activity 
    as a function of exercise intensity, 
    simulations of 
    an exercise program of three sessions/week, 60 minutes/session in a time window of 5 years, with $u$ varying from $0\%$ to $70\%$ were performed to span the full range of light, moderate and vigorous intensity.
    In addition, the same training program was simulated on a time window of 20 years considering moderate and vigorous intensities ($u=50\%$, $60\%$, and $70\%$).  
    \subsubsection*{Simulations of a regular exercise program following the WHO preventive recommendations}

    The WHO addressed the problem of preventing cardiovascular and chronic diseases through exercise by developing the following, equivalent recommendations 
    \cite{bull2020world}:
     
     \begin{itemize}
            \item[-]at least 75 minutes/week  of vigorous-intensity aerobic physical activity;
             \item[-]at least 150 minutes/week  of moderate-intensity aerobic physical activity.
        \end{itemize}
Moreover, as suggested by Boonpor et al. \cite{boonpor2023dose}, the benefit of moderate-intensity aerobic physical activity tends to saturate at about 400 minutes/week. To assess the ability of the proposed model to
reflect the expected benefits of the WHO recommendations
, the two suggested training programs were 
simulated by distributing the weekly duration in three sessions, each lasting 25 and 50 minutes, respectively. The variable $u$ was set equal to $75\%$ for vigorous intensity and equal to the $50\%$ for moderate intensity \cite{roy2007dynamic}.  
    In addition, to assess if the proposed model is able to replicate the expected saturation in benefit for weekly duration of moderate-intensity exercise equal to or higher than 400 minutes/week \cite{boonpor2023dose},
    simulations were performed for a total duration of 300, 350, 400, 450 and 500 minutes/week, distributed in three exercise sessions per week, by setting $u=50\%$.
    \subsubsection*{Simulations of a discontinued exercise program}

The literature provides evidence that the benefits 
of exercise persist after the discontinuation of the intervention \cite{lindstrom2006sustained,li2008long}. 
Specifically, in the work by Lindstr\"om et al. \cite{lindstrom2006sustained} (in the following referred to as FDPS) adults with impaired glucose tolerance were involved in a four-year lifestyle intervention program that included moderately intense physical activity for a minimum of 30 minutes per day. 
Participants were followed for three more years after the intervention.
Results outlined that the incidence rates of T2D and the glycemia values at the end of the intervention and in the post-intervention period varied only slightly, suggesting that beneficial lifestyle effects were maintained after the discontinuation of the intervention.
Similar findings are described in Li et al. \cite{li2008long,pan1997effects}(in the following referred to as CDQDPS), in which adults with impaired glucose tolerance were involved in a 20-year follow up, with lifestyle intervention (moderate intensity exercise, 2 sessions/day, 20 minutes/session) discontinued at the sixth year and benefits in terms of reduced T2D incidence were observed up to 14 years after the intervention.
Scenarios similar to the ones described by the two studies were simulated, 
    specifically:
    \begin{itemize}
            \item[-] for FDPS: moderate intensity exercise was simulated
            at three different values of $u$ ($30\%$, $40\%$, $50\%$), 
            with daily sessions 
            of 30 minutes. The exercise was interrupted at the fourth year, with a simulation horizon of seven years.
            \item[-] for CDQDPS: moderate intensity exercise was simulated 
            at three different values of $u$ ($30\%$, $40\%, 50\%$),  
            with two exercise sessions per day, each lasting twenty minutes,
            as in suggested in Pan et al.\cite{pan1997effects}. The exercise was interrupted at the sixth year, with a simulation horizon of twenty years.
        \end{itemize}
    
\begin{table}[!ht]
    \centering
    \begin{tcolorbox}[colback=white!10, colframe=white!100!black, sharp corners, boxrule=1pt, width=\textwidth]
        \centering
        \footnotesize 
        \begin{tabular}{p{1.5cm}@{}l*{12}{c}@{}}
            \toprule
             & & & $G_{10-year}$ & $G_{15-year}$ & $G_{20-year}$ & $I_{10-year}$ & $I_{15-year}$ & $I_{20-year}$ \\ 
             && & (mg/dl) & (mg/dl) & (mg/dl) & ($\mu$U/ml) & ($\mu$U/ml) & ($\mu$U/ml) \\
            \midrule 
            \multirow{2}{*}{$u=50\%$} & $\tau_{SI}=150$ & & 128 & $>>$150 & $>>$150 & 29 & $<3$ & $<3$ \\
            & $\tau_{SI}=180$&  & 116 & 105 & 101 & 33 & 38 & 40 \\
            \addlinespace
            \midrule 
            \multirow{2}{*}{$u=60\%$} &$\tau_{SI}=150$ &  & 117 & 108 & 101 & 33 & 36 & 39 \\
            & $\tau_{SI}=180$& & 111 & 101 & 100 & 35 & 39 & 40 \\
            \addlinespace
            \midrule 
            \multirow{2}{*}{$u=70\%$} &$\tau_{SI}=150$ & & 112 & 102 & 100 & 35 & 39 & 40 \\
            &$\tau_{SI}=180$& & 108 & 101 & 100 & 36 & 39 & 40 \\
            \bottomrule
        \end{tabular}
        \caption{{Basal} glucose and insulin concentration {observed} over a 20-year {horizon} for {$u=50\%$, $60\%$, and $70\%$} and {$\tau_{SI}=150$ and $180$ days}.}
    \label{tab:GlucoseTrend1}
    \end{tcolorbox}
    
\end{table}
\begin{table}[!h]
\centering

\begin{tcolorbox}[colback=white!10, colframe=white!100!black, sharp corners, boxrule=1pt,width=\textwidth]
\begin{tabular}{@{}l*{7}{c}l*{7}{c}@{}}
FDPS & &$G_{4\text{th-year}}$ &  &$G_{7\text{th-year}}$ && CDQDPS & &$G_{6\text{th-year}}$ &  &$G_{20\text{th-year}}$ \\ 
&&(mg/dl)& & (mg/dl) &&&&(mg/dl)& & (mg/dl) \\
\midrule 
$u=30\%$  &$\tau_{SI}=150$& 131 & &$>>150$ && $u=30\%$ &$\tau_{SI}=150$& 133 & &$>>150$ \\ 
& $\tau_{SI}=180$ &126& &136 &&& $\tau_{SI}=180$ &125& &$>>150$ \\
\addlinespace
$u=40\%$ &$\tau_{SI}=150$& 127& &$>>150$ && $u=40\%$ &$\tau_{SI}=150$& 124& &123 \\ 
& $\tau_{SI}=180$ &123 & &126 &&& $\tau_{SI}=180$ &119 & &101 \\
\addlinespace
$u=50\%$  &$\tau_{SI}=150$ & 124 & & 128 && $u=50\%$ &$\tau_{SI}=150$ & 119 & &101\\
& $\tau_{SI}=180$ &120& &121 &&& $\tau_{SI}=180$ &115& &100 \\
\addlinespace
\end{tabular}

\caption{Basal glucose concentration ($G$) observed in the simulations of the FDPS and CDQDPS programs, with $\tau_{SI}=150$ and $\tau_{SI}=180$ days and with $u$ in the moderate-intensity range.}
\label{tab:CombinedTable}
\end{tcolorbox}

\end{table}

\section*{Data availability}
The datasets generated  during the current study may be made available to qualified researchers on reasonable request to the corresponding author.

\section*{Code availability}
The underlying code for this study is not publicly available but may be made available to qualified researchers on reasonable request to the corresponding author.

\section*{ Acknowledgments}
{This work was supported in part by the European Union through the Project PRAESIIDIUM “Physics Informed Machine Learning-Based Prediction and Reversion of Impaired Fasting Glucose Management" (call HORIZON-HLTH-2022-STAYHLTH-02) under Grant 101095672. Views and opinions expressed are however those of the authors only and do not necessarily reflect those of the European Union. The European Union can not be held responsible for them.\\
This work was carried out within the Italian National Ph.D. Program in Autonomous Systems (DAuSy), coordinated by Politecnico of Bari, Italy.
}

\section*{Author Contributions}
{PFDP designed the study, developed the code, performed the simulations, wrote and revised the manuscript. AB participated in the design of the study, contributed to the code development, revised the manuscript. FD participated in the design of the study, supervised the development of the work, revised the manuscript. KK contributed to the development of the work and to the interpretation of results. PP participated in the design of the study and revised the manuscript. AP designed the study, supervised the development of the work, contributed to the interpretation of results, wrote and revised the manuscript. {All authors read and approved the final manuscript.}}

\subsection*{Competing interests}
NA

\bibliographystyle{naturemag}
\bibliography{model}
\end{document}